 \documentclass[aip,jap,preprint,floatfix,showpacs,amsmath,amssymb]{revtex4-1}
\usepackage{natbib}
\usepackage{float}
\usepackage[dvips,final]{graphicx}
\usepackage{color}
\usepackage{amsmath}
\usepackage{mathrsfs} 

\usepackage{graphicx, subfigure}
\usepackage{pslatex}
\usepackage{relsize}
\usepackage{bm}
\usepackage{xspace} 

\usepackage{xcolor}
\usepackage{soul}

\def\ba{\begin{eqnarray}}
\def\ea{\end{eqnarray}}
\def\be{\begin{equation}}
\def\ee{\end{equation}}

\bibpunct{[}{]}{,}{n}{}{} 

\begin{document}
\title{Magnetic ground states for bent nanotubes}

\author{D. Mancilla-Almonacid}
\address{Departamento de F\'isica, CEDENNA, Universidad de Santiago de Chile, Avenida Ecuador 3493, Santiago, Chile}
\author{M. A. Castro}
\address{Departamento de F\'isica, CEDENNA, Universidad de Santiago de Chile, Avenida Ecuador 3493, Santiago, Chile}
\author{J. M. Fonseca}
\address{Departamento de F\'isica, Universidade Federal de Vi\c cosa, Avenida Peter Henry Rolfs s/n, 36570-000 Vi\c cosa, MG, Brazil}
\author{D. Altbir}
\address{Departamento de F\'isica, CEDENNA, Universidad de Santiago de Chile, Avenida Ecuador 3493, Santiago, Chile}
\author{S. Allende}
\address{Departamento de F\'isica, CEDENNA, Universidad de Santiago de Chile, Avenida Ecuador 3493, Santiago, Chile}
\author{V. L. Carvalho-Santos}
\address{Departamento de F\'isica, Universidade Federal de Vi\c cosa, Avenida Peter Henry Rolfs s/n, 36570-000 Vi\c cosa, MG, Brazil}
\email{vagson.carvalho@usachcl}
\begin{abstract}

Magnetic nanotubes have been widely studied because they are promising candidates to be part in devices based on spintronic and magnonic technologies. However, the experimental techniques used to prepare these elements could not guarantee to have perfect nanostructures. Therefore,  some geometric imperfections can appear. In this direction,   the bent of a nanotube could play an essential role in the magnetic properties of a device. In this work, we analyze the influence of curvature on the magnetic properties of a bent nanotube, a topic scarcely studied, and that can have a strong impact on applications.
\end{abstract}

\maketitle

\section{Introduction}

Nanomagnetism is an area of great interest for basic research and technological applications. Indeed, the possibility of using nanomagnets in random access memory, data storage \cite{Paper-1}, spintronic, and magnonic devices \cite{paper-3} has made magnetism at the nano-scale a very prominent topic. New technological developments made it possible the production and characterization of nanomagnets with several shapes and sizes, e.g., rolled-up magnetic membranes \cite{paper-4}, paraboloidal caps \cite{KaiLiu-paper}, modulated nanomagnets \cite{Knielsh-paper} and cylindrical nanowires (NW) and nanotubes (NT) \cite{experimental-techniques}. The fabrication of magnetic nanostructures with different shapes has promoted the study on how geometry influences their magnetic properties, highlighting the influence of the curvature on the properties of nanosized elements in 1, 2, and 3 dimensions. Then,  the interplay between geometry and magnetism in several and different contexts is very relevant \cite{Gaididei-PRL-2014,Review-Curvature,Review-3DNanomagnetism,Landeros-JAP-2010,Hertel-Spin-2013,Pylypvskyi-SRep-2016,Smiljan-JAP-2017,Otalora-PRL-2017}. 

Due to their relatively simple theoretical description and easy fabrication process, cylindrical nanomagnets have been widely studied. Several works are devoted to describing statical and dynamical properties of the magnetization in cylindrical nanodots \cite{dots1}, nanorings \cite{rings1}, NW, and NT \cite{wires}. In particular, cylindrical nanowires and nanotubes can be produced in large quantities \cite{experimental-techniques,Knielsh} and are prominent candidates to compose units of information for non-volatile memories, logic gates \cite{Jaworowicz-Nanotechnology-2009,Allwood-Science-2005}, drug delivery containers \cite{Son-JChem} and racetrack memory devices \cite{Parkin-Science-2008}. Theoretical \cite{theoretical-studies,Yan-paper,theoretical-studies2,theoretical-studies3} and experimental \cite{experimental-techniques,Knielsh} results have shown the appearing of interesting phenomena related to the interplay between the cylindrical geometry and the domain wall dynamics along NW and NT. For instance, the absence of a Walker field \cite{Walker-paper} in the domain wall (DW) motion along cylindrical NW is very relevant. Indeed, unlike the described dynamics of a DW displacing along in a magnetic nanostripe \cite{Mougin-paper}, the DW propagation along a cylindrical NW presents a velocity that varies linearly with the magnetic field \cite{Yan-paper}.  

Nevertheless, if cylindrical NW or NT are bent, the lost of the cylindrical symmetry results in interesting magnetic properties. For example, Kravchuk{\it et al.} \cite{Kravchuk-wire-paper} showed that a domain wall can be pinned even when it is propagating across a bent region of a cylindrical wire. Also, Moreno {\it et al.} \cite{Moreno-PRB-2017} concluded that the oscillatory motion related to the Walker breakdown is recovered when a domain wall is displacing along a bent nanowire with a cylindrical cross-section. 

In this context, the discussion on the properties of bent nanotubes (BNT) with a cylindrical cross-section is particularly interesting because the lack of a magnetic core can make flux-closure magnetization configurations more favorable than single-domain states \cite{Escrig-JMMM-2007}. However, since the magnetic ground state of a nanoparticle depends on its geometry, the minimum energy magnetization configuration of a bent NT with cylindrical cross-section must present differences when compared with a straight tube with the same geometrical parameters. For instance, it was shown that chiral effects coming from the variable curvature of toroidal NTs are responsible for the nucleation of topological solitons with positive (negative) winding numbers in regions with positive (negative) curvature of a magnetic nanoparticle \cite{Smiljan-JAP-2017}. Due to the interesting variety of phenomena that appear in curved systems, in this work we calculate the magnetic energies of different magnetization configurations of a BNT and obtain a phase diagram describing the magnetization configuration presenting the minimum energy. From the obtained phase diagram, we described the critical thickness as a function of the curvature and length for which the flux-closure magnetization, here called vortex state, loses its stability. The analyzed magnetization patterns are analogous to magnetic states obtained in straight cylindrical nanomagnets \cite{Escrig-JMMM-2007}, that is; \textit{i}) in-surface state ($IS$); and \textit{ii}) vortex state ($V$). 

This work is organized as follows. In Section \ref{ThModel} we present the adopted theoretical model to describe the nanotube and to calculate the magnetic energy; Section \ref{ThResults} contains results and discussions;  finally, in Section \ref{ThConclusions} the conclusions and prospects are presented.

\section{Theoretical model}\label{ThModel}

Aiming to obtain the magnetic state diagram of a bent cylindrical NT, we start calculating the magnetic energy of different magnetization configurations, similar to those that appear in straight cylindrical NT. By considering polycrystalline materials we neglect anisotropy contributions to the energy and then the total magnetic energy will be given by $E_t=E_x+E_m$, where $E_t$, $E_x$ and $E_m$ are the total, exchange, and magnetostatic energies, respectively. The energy calculations are performed using the micromagnetic theory, in which a magnetic element is described as a continuous medium whose magnetic state is defined by a vector field that is a function of the position inside the magnet. 

The exchange energy can be calculated from a continuous approximation of the Heisenberg model, given by
\be\label{exen}
E_x=A\int_V[( \boldsymbol{\nabla}m_x)^2+( \boldsymbol{\nabla}m_y)^2+( \boldsymbol{\nabla}m_z)^2]\,dV\,,
\ee

\noindent
where $A$ is the stiffness constant, $\boldsymbol{ \nabla}$ represents the gradient operator in a curvilinear basis, and $\mathbf{m}=\mathbf{M}/M_s$, with $\mathbf{M}$ the magnetization vector, and $M_s$ the saturation magnetization. 

The magnetostatic energy can be determined from the interaction between the magnetization and the demagnetizing field generated by the magnetization itself; that is,
\ba\label{dipen}
E_{m}=-\frac{\mu_o M_s}{2}\int_V\mathbf{H}_d\cdot\mathbf{m}\,dV\,,
\ea

\noindent
where $\mu_0$ is the magnetic permeability. The vector field $\mathbf{H}_d$ is the demagnetizing field, which can be determined by $\mathbf{H}_d=-\boldsymbol{\nabla}\Phi$, where $\Phi$ is the magnetostatic potential that can be obtained from the Laplace equation $\nabla^2 \Phi=0$. In the absence of electric currents, the magnetostatic potential comes from surface, $\sigma$, and volumetric, $\chi$, magnetic charges defined as $\sigma=\mathbf{m}\cdot\mathbf{n}$ and $\chi=- \boldsymbol {\nabla}\cdot\mathbf{m}$, where  $\mathbf{n}$ is the normal vector pointing outward the surface of the structure. The formal solution of the Laplace equation is given by
\begin{eqnarray}
\Phi(\mathbf{r}')\equiv\Phi=\frac{M_s}{4\pi}\left[\int_S\frac{\sigma}{\mathcal{R}}dS+\int_V\frac{\chi}{\mathcal{R}}dV\right]\,,
\end{eqnarray}

\noindent
where $\mathcal{R}\equiv|\mathbf{r}\,'-\mathbf{r}|$ is the distance between two points inside the magnetic body.

\begin{figure}
\includegraphics[scale=0.55]{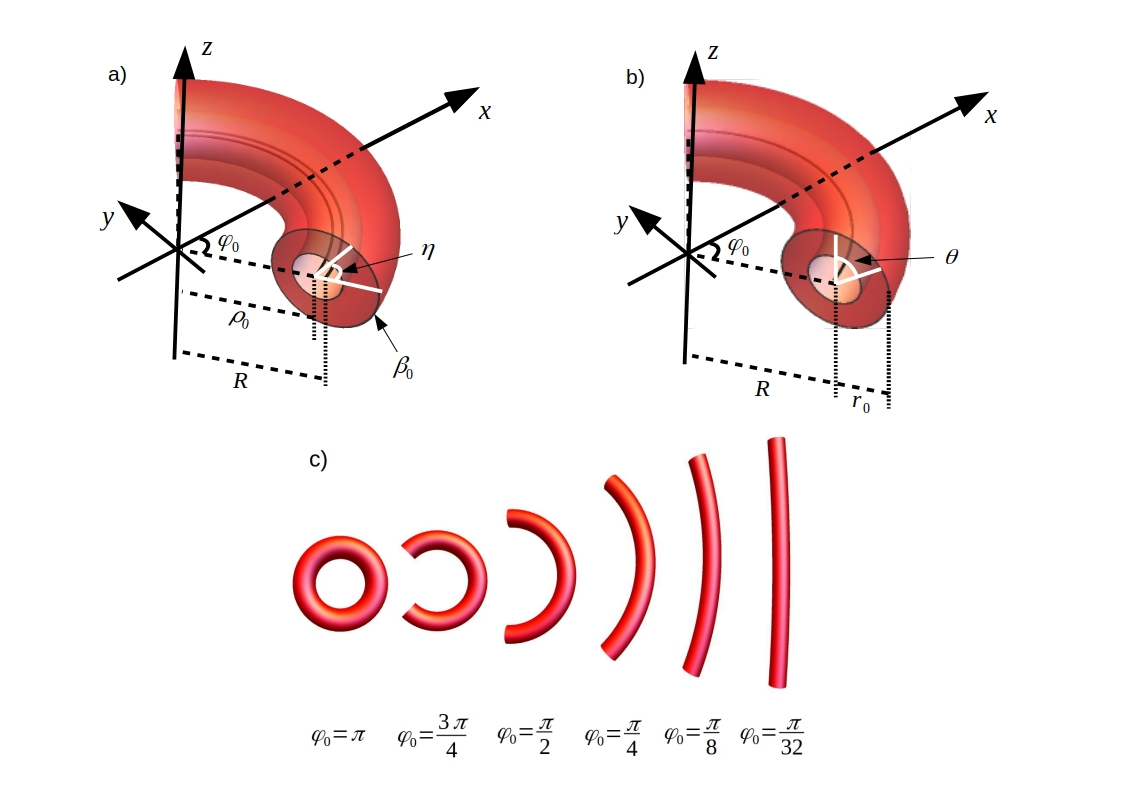}\caption{Schematic representation of a bent nanotube. (a) Bent nanotube represented in the toroidal coordinate system. (b) Bent nanotube represented in the spherical-like coordinate system. (c) Schematic representation of a bent nanotube for† different † $\varphi_0$ values.}\label{CoordinateSystem}
\end{figure}

In our study, the geometrical description of a curved magnetic NT is given considering it as a toroidal section parametrized as \cite{Morse-Book}
\begin{eqnarray}\label{CoordSystem}
\mathbf{\mathbf{r}}=\hat{\textbf{x}}\frac{\rho\sinh\beta\cos\varphi}{\cosh\beta-\cos\eta}+\hat{\textbf{y}}\frac{\rho\sinh\beta\sin\varphi}{\cosh\beta-\cos\eta}+\hat{\textbf{z}}\frac{\rho\sin\eta}{\cosh\beta-\cos\eta}\,,
\end{eqnarray}

\noindent
where $\rho$ is a constant that defines the radius of a circle in the plane $z=0$ when $\beta\rightarrow\infty$. Here $\beta\in[\beta_0,\infty)$ determines the tube external radius ($\beta_0$ describes the external surface), $\varphi\in[-\varphi_0,\varphi_0]$ and $\eta\in[0,2\pi]$ are  the azimuthal and  poloidal angles, respectively. Fig. \ref{CoordinateSystem}(a) shows the geometrical parameters describing a BNT according to the parametrization given in Eq. \eqref{CoordSystem}. It is worth to notice that a nanotube with constant thickness (measured from the external to the internal border of the tube) is not obtained from varying $\beta$ from $\beta_0$ to $\beta_1$. Indeed, if we simply vary $\beta$, we would obtain a nanotube with a variable thickness along the poloidal angle [see Fig. \ref{CoordinateSystem}(a)]. In this context, and to determine the magnetostatic energy, we use the superposition principle; that is, we consider that the nanotube is formed by two nanowires with opposite magnetizations. The normal vector pointing outward the external surface of the tube is given by $\mathbf{n}=-\boldsymbol{\hat\beta}=-F(\beta,\eta)\mathbf{\hat{R}}-G(\beta,\eta)\mathbf{\hat{z}}\,,$ where
\begin{eqnarray}\label{ConvFactors}
F(\beta,\eta)\equiv F=\frac{1-\cosh\beta\cos\eta}{{\cosh\beta-\cos\eta}}\,,\nonumber\\
G(\beta,\eta)\equiv G=-\frac{\sinh\beta\sin\eta}{\cosh\beta-\cos\eta}\,,
\end{eqnarray}

\noindent
and $\mathbf{\hat{R}}=\mathbf{\hat{x}}\cos\varphi+\mathbf{\hat{y}}\sin\varphi$ is the radial vector pointing along the $xy$-plane, outward the bent. Finally, the components of a vector field in their cartesian components can be promptly obtained from the components of a vector field ($\mathbf{H}$) written in terms of Eq. (\ref{CoordSystem}) ($\beta, \eta, \varphi$) by using the transformation
\begin{eqnarray}\label{Conversion}
\left(\begin{array}{c}
H_x\\H_y\\H_z
\end{array}\right)=\left(\begin{array}{ccc}
F\cos\varphi & G\cos\varphi & -\sin\varphi\\F\sin\varphi & G\sin\varphi & \cos\varphi\\G & -F & 0
\end{array}\right)\left(\begin{array}{c}
H_\beta\\H_\eta\\H_\varphi
\end{array}\right).
\end{eqnarray}

Additionally, the description of a BNT can be obtained using a more intuitive parametrization describing a toroidal section; that is, by using the simple toroidal coordinates
\ba\label{CoordSystem2}
{\mathbf{r}}=\mathbf{\hat{R}}\,(R+r\sin\theta)+\mathbf{\hat{z}}\,r\cos\theta\,,
\ea

\noindent
where $r$ and $R$ are the poloidal and ``toroidal'' radii [see Fig.~\ref{CoordinateSystem}(b)], respectively, and $\theta\in[0,2\pi]$ is equivalent to the polar angle of the spherical coordinate system. The external border of the tube is described by a bent cylindrical wire (BCW) with radius $r_0=R(\cosh\beta_0)^{-1}$ and $\rho_0=(R^2-r_0^2)^{1/2}$, with a magnetization equal to $\mathbf M$, as long as the internal border is described by a BCW with radius $r_1=R(\cosh\beta_1)^{-1}$ and $\rho_1=(R^2-r_1^2)^{1/2}$, with a magnetization equal to $-\mathbf M$, [see Fig.~\ref{CoordinateSystem}(b)]. The tube length is defined as $L=2\varphi_0 R$ in such a way that the BNT with greater curvature, described by a torus, is obtained for $\varphi_0=\pi$, and an almost straight BNT is described by $R\rightarrow\infty$ and $\varphi_0 \rightarrow 0$. Bent nanotubes for different $\varphi_0$ values and fixed $L$ are shown in Fig.~\ref{CoordinateSystem}(c).

The calculation of the magnetostatic potential is, in general, very hard from the mathematical side. A possible way to perform it is by expanding the inverse of the distance in an infinite series using Green's functions. For the considered geometry, the inverse of the distance can be written as a function of the coordinate system given by Eq. (\ref{CoordSystem}), allowing us to write\cite{Gradshtein-Book}
\begin{eqnarray}\label{GreenFunc}
\frac{1}{\mathcal{R}}=\frac{\sqrt{\Upsilon'\, \Upsilon}}{\pi\,\rho}\sum_{k,n=0}^{\infty}(-1)^k\varepsilon_n\varepsilon_k\cos [n(\eta'-\eta)]\cos [k(\varphi'-\varphi)]\nonumber\\\times\frac{\Gamma(n-k+1/2)}{\Gamma(n+k+1/2)}P_{n-1/2}^k(\cosh\beta_<)Q_{n-1/2}^k(\cosh\beta_>)\,,\,\,\,
\end{eqnarray}

\noindent
where $\Upsilon'=\cosh\beta'-\cos\eta'$, $\Upsilon=\cosh\beta-\cos\eta$, $\varepsilon_{k}=(2-\delta_{k,0})$, $\varepsilon_{n}=(2-\delta_{n,0})$, $\Gamma(x)$ is the gamma function, $P_{n-1/2}^k(\cosh\beta)$ and $Q_{n-1/2}^k(\cosh\beta)$ are the Legendre functions of half-integer order (also known as toroidal harmonics)  of the first and second kind, respectively. From now on, we will adopt the following notation to represent the toroidal harmonics
\begin{eqnarray}
P^{\tau}_{\nu}(\cosh x)=P^{\tau\,\,\,\,x}_{\nu}\,\,\,\,\,\,\,\text{and}\,\,\,\,\,\,\,
Q^{\tau}_{\nu}(\cosh x)=Q^{\tau\,\,\,\,x}_{\nu}\,.
\end{eqnarray}

\begin{figure}
\includegraphics[scale=0.4]{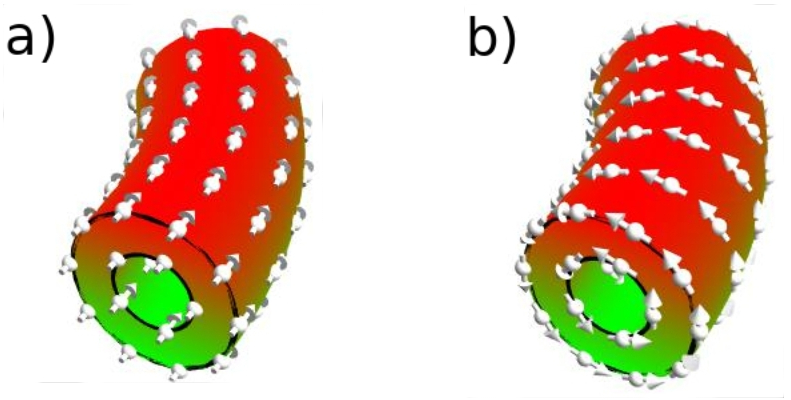}\caption{Schematic representation of the magnetic configurations for a bent nanotube. (a) $IS$ state and (b) $V$ state.}\label{MagConfig}
\end{figure}

To obtain the magnetic ground state for a bent nanotube as a function of its geometry, we will calculate the magnetic energy associated with two different magnetization configurations lying on a BNT (see Fig. \ref{MagConfig}): i) an in-surface state ($\text{IS}$), in which the magnetic moments point along the azimuthal direction of the nanotube, described as $\mathbf{m}_{{\text{IS}}}=\pm\boldsymbol{\hat{\varphi}}$; and ii) a vortex state ($V$), in which the magnetic moments turn around the poloidal angle of the nanotube, forming a flux-closure structure, represented by $\mathbf{m}_{{{V}}}=\pm\boldsymbol{\hat{\theta}}$. In the above referenced representations, the plus and minus signs refer to the magnetization of the BNW with radii $r_0$ and $r_1$, respectively. 


\section{Analytical results}\label{ThResults}

Using the presented model, analytical expressions for the magnetostatic and exchange energies can be obtained. At first place, we calculate the exchange energy of the considered magnetization configurations. Since the exchange interaction is of short range, it can be easier obtained by describing the BNT in the frame of Eq. \eqref{CoordSystem2}. In this context, Eq. (\ref{exen}) can be rewritten as
\begin{eqnarray}\label{exen1}
E_{x}=A\int_{r_1}^{r_0}dr\int_S [( \boldsymbol{\nabla}m_x)^2+( \boldsymbol{\nabla}m_y)^2+( \boldsymbol{\nabla}m_z)^2] dS\,,
\end{eqnarray}

\noindent
where the gradient operator and the surface element must be written in terms of the curvilinear basis described in Eq. (\ref{CoordSystem2}), that is, $
\boldsymbol{\nabla}=r^{-1}\,\boldsymbol{\hat{\theta}}\,{\partial_\theta}+(R+r\sin\theta)^{-1}\,\boldsymbol{\hat{\varphi}}\,{\partial_\varphi}$, $dS=r(R+r\sin\theta)d\theta d\varphi$, and the magnetization $\mathbf{M}\equiv M_s\mathbf{m}$ can be parametrized in a spherical coordinate system ($r,\theta,\phi$) on the basis of cartesian coordinates, that is, $\mathbf{m}=\hat{\mathbf{x}}\cos\phi\sin\theta+\hat{\mathbf{y}}\sin\phi\sin\theta+\hat{\mathbf{z}}\cos\theta$. 
 
The substitution of the magnetization profile describing the $V$ state in Eq. (\ref{exen1}) yields  the corresponding exchange energy for a $V$ state in a BNT ($E_{x_{V}}$)
\ba\label{exEneV}
E_{x_{V}}=2\pi A L\left[\ln\left(\frac{r_0}{r_1}\frac{\sqrt{R^2-r_0^2}+R}{\sqrt{R^2-r_1^2}+R}\right)+\frac{1}{R}\left(\sqrt{R^2-r_ 1^2}-\sqrt{R^2-r_0^2}\right)\right]\,.
\ea

The exchange energy in an equivalent straight tube, $E_{{x}_{c}}$, is $E_{x_{c}}=2\pi A L\,\ln({r_0}/{r_1})$ \cite{Escrig-JMMM-2007}. We observe that 
$E_{x_{c}} \leqslant E_{x_{V}}$, and the equality is valid only in the limit $R\rightarrow\infty$ and $\varphi_0 \rightarrow 0$. This behavior occurs because in a BNT, due to the curvature, there is an extra deviation of the magnetic moments from the parallel alignment (observed in a $V$ state in a straight NT), increasing accordingly the exchange energy.

Eq. \eqref{exen1} can be also used to obtain the exchange energy of the $IS$ state, giving
\ba\label{exEneIS}
E_{x_{IS}}=\frac{2\pi AL}{R}\left[\sqrt{R^2-r_1^2}-\sqrt{R^2-r_0^2}\right]\,,
\ea

\noindent
which vanishes in the limit $R\rightarrow\infty$ due to the parallel alignment of the magnetic moments in this limit. It should also be noted  that for this state the larger the curvature, the larger the exchange energy.

Aiming to obtain the magnetostatic energy of the considered magnetization configurations, we calculate the magnetostatic potential of each magnetization profile and their respective demagnetizing fields. As explained before, we use the superposition principle and assume that the nanotube can be considered as two nanowires with different radii ($r_0$ and $r_1$) and opposite magnetizations.

Despite the absence of surface magnetic charges, the $V$ state presents a dipolar energy related to volumetric magnetic charges given by $\chi=-\frac{1}{r_0}\frac{\sin \eta}{\sinh \beta} \frac{\sqrt{\cosh (\beta)-\cos (\eta )}}{  \sqrt{\cosh (\beta -2 \text{$\beta_0 $})-\cos (\eta)}}$. Thus, the magnetostatic potential of the $V$ state of a BCW, $\tilde{\Phi}_{V}$, with radius $r_0 $ ($\cosh \beta_0=R/r_0$) and $\rho_0=(R^2-r_0^2)^{1/2}$
\ba\label{PotVortex2}
\tilde{\Phi}_{V}(\rho_0,\beta_0)=- \frac{\rho_0 ^2}{r_0}\frac{M_s}{4\pi^2 }  \sqrt{\Upsilon'} \sum_{k,n}^{\infty} (-1)^k \epsilon_k \epsilon_n \sin{(n\eta')} h_k(\varphi')  \mathcal{V}^k_n(\beta',\beta_0)\,, 
\ea
where $h_k(\varphi')=k^{-1}[2\cos ( k\varphi')\sin (k\varphi_0)]$ and
\ba
\mathcal{V}^k_n(\beta',\beta_0)=&[ Q^{-k\,\,\,\,\beta'}_{n-1/2}  \int_{\beta_0}^{\beta'}     P^{k\,\,\,\,\beta}_{n-1/2}\int_0^{2\pi} \frac{\sin{(n\eta)}\sin \eta}{ (\cosh (\beta -2 \text{$\beta_0 $})-\cos (\eta ))^{1/2}(\cosh (\beta)-\cos (\eta ))^2}  d\eta d\beta \\
+& P^{k\,\,\,\,\beta'}_{n-1/2}\int_{\beta'}^{\infty}  Q^{-k\,\,\,\,\beta}_{n-1/2}\int_0^{2\pi} \frac{\sin{(n\eta)}\sin \eta}{ (\cosh (\beta -2 \text{$\beta_0 $})-\cos (\eta))^{1/2}(\cosh (\beta)-\cos (\eta))^2}  d\eta  d\beta].\nonumber
\ea
Then, the magnetostatic potential for the vortex state in the BNT is
\ba\label{magtotal}
\Phi_V= \tilde{\Phi}_{V}(\rho_0,\beta_0) - \tilde{\Phi}_{V}(\rho_1,\beta_1).
\ea

For details on the magnetostatic potential calculation, see Appendix\ref{Apendice}. For the $IS$ configuration, $\nabla\cdot\mathbf{m}=0$, and then, only surface magnetic charges play a role in the magnetostatic energy calculations. The surface magnetostatic contribution comes from the borders of the tube, located at $-\varphi_0$ ($\sigma=-1$) and $\varphi_0$ ($\sigma=1$). In this case, after some algebraic manipulation, the magnetostatic potential of the $IS$ state of a BCW, $\tilde{\Phi}_{{\text{IS}}}$, with radius $r_0 $ ($\cosh \beta_0=R/r_0$) and $\rho_0=(R^2-r_0^2)^{1/2}$ is 

\begin{eqnarray}\label{PotSX1}
\tilde{\Phi}_{{\text{IS}}}(\rho_0,\beta_0)={\frac{2\sqrt{2}\,M_s}{\pi^2}\,\rho_0\sqrt{\Upsilon'}}\sum_{k,n=0}^\infty (-1)^k\varepsilon_n\varepsilon_k\,\cos (n\eta')\,\,g_k(\varphi') \mathcal{W}^k_{n}(\beta',\beta_0)\,\,\,\,\,,
\end{eqnarray}

\noindent
where $g_k(\varphi')=\sin (k\varphi_0)\,\sin (k\varphi')$ and
\ba
\mathcal{W}^k_n(\beta',\beta_0)=\left[Q^{-k\,\,\,\,\beta'}_{n-1/2}\int_{\beta_0}^{\beta'}\frac{\,P^{k\,\,\,\,\beta}_{n-1/2}Q^{1\,\,\,\,\beta}_{n-1/2}}{\sinh\beta}\,d\beta+P^{k\,\,\,\,\beta'}_{n-1/2}\int_{\beta'}^{\infty}\frac{Q^{-k\,\,\,\,\beta}_{n-1/2}Q^{1\,\,\,\,\beta}_{n-1/2}}{\sinh\beta}\,d\beta\right]\,.\nonumber
\ea

Then, the magnetostatic potential for the in-surface state in the BNT is
\ba\label{magtotal}
\Phi_{IS}= \tilde{\Phi}_{IS}(\rho_0,\beta_0) - \tilde{\Phi}_{IS}(\rho_1,\beta_1).
\ea

Finally, we are in position to determine the magnetostatic energy of each considered magnetization configuration. At first place, we determine the magnetostatic energy of the $V$ state. From Eq. \eqref{PotVortex2} we can determine the $\hat{\theta}$ component of the demagnetizing field and then, after some algebra, the magnetic energy of the $V$ configuration is evaluated as 

\begin{align}
\tilde{E}_{m_{V}}(\rho_0,\beta_0)=\frac{\mu_0 M_s\rho_0^2}{2} \int_{0}^{2\pi} \int_{0}^{2\pi}\int_{\beta_0}^{\infty} \left(\frac{\partial \tilde{\Phi}_{V}}{\partial \beta'}c_1+\frac{\partial \tilde{\Phi}_{V}}{\partial \eta'}c_2\right) \frac{ \sinh (\beta' )}{[\cosh (\beta' )-\cos (\eta' )]^2} d\beta' d\eta' d\varphi',
\label{dipEneV}
\end{align}
where
\begin{align}
&c_1=
-\frac{\sin (\eta' ) \sinh (\beta' -\text{$\beta_0 $})}{[(\cosh (\beta' )-(\cos (\eta' )) (\cosh (\beta' -2 \text{$\beta_0 $})-(\cos (\eta' ))]^{1/2}},\nonumber\\
&c_2=\frac{\cos (\eta' ) \cosh (\beta' -\text{$\beta_0 $})-\cosh (\text{$\beta_0 $})}{[(\cosh (\beta' )-(\cos (\eta' )) (\cosh (\beta' -2 \text{$\beta_0$})-(\cos (\eta' ))]^{1/2}}.
\end{align}

\noindent
Then, the magnetostatic energy for the vortex state in the BNT is
\ba\label{magtotal}
{E}_{m_{V}}= \tilde{E}_{m_{V}}(\rho_0,\beta_0)-\tilde{E}_{m_{V}}(\rho_1,\beta_1) \geq 0.
\ea

The energy of the $IS$ state can be obtained in an analogous way from the substitution of Eq.~\eqref{PotSX1} into Eq.~\eqref{dipen}, resulting in
\ba\label{dipEneIS}
\tilde{E}_{m_{\text{IS}}}(\rho_0,\beta_0)=\frac{16\mu_0 M_s^2 \rho_0^3}{\pi^2}  \sum_{k,n=0}^{\infty}(-1)^{k}\varepsilon_k\varepsilon_n\,\sin^{2}(k\varphi_0)\,\int_{\beta_0}^{\infty}\mathcal{W}^{k}_{n}(\beta',\beta_0)\frac{Q^{1\,\,\,\,\beta'}_{n-1/2}}{\sinh\beta'}\,d\beta'.
\ea
Then, the magnetostatic energy for the in-surface state in the BNT is
\ba\label{magtotal}
{E}_{m_{IS}}= \tilde{E}_{m_{IS}}(\rho_0,\beta_0)-\tilde{E}_{m_{IS}}(\rho_1,\beta_1) \geq 0.
\ea

\section{Results and Discussion}

\begin{figure}
\includegraphics[scale=0.7]{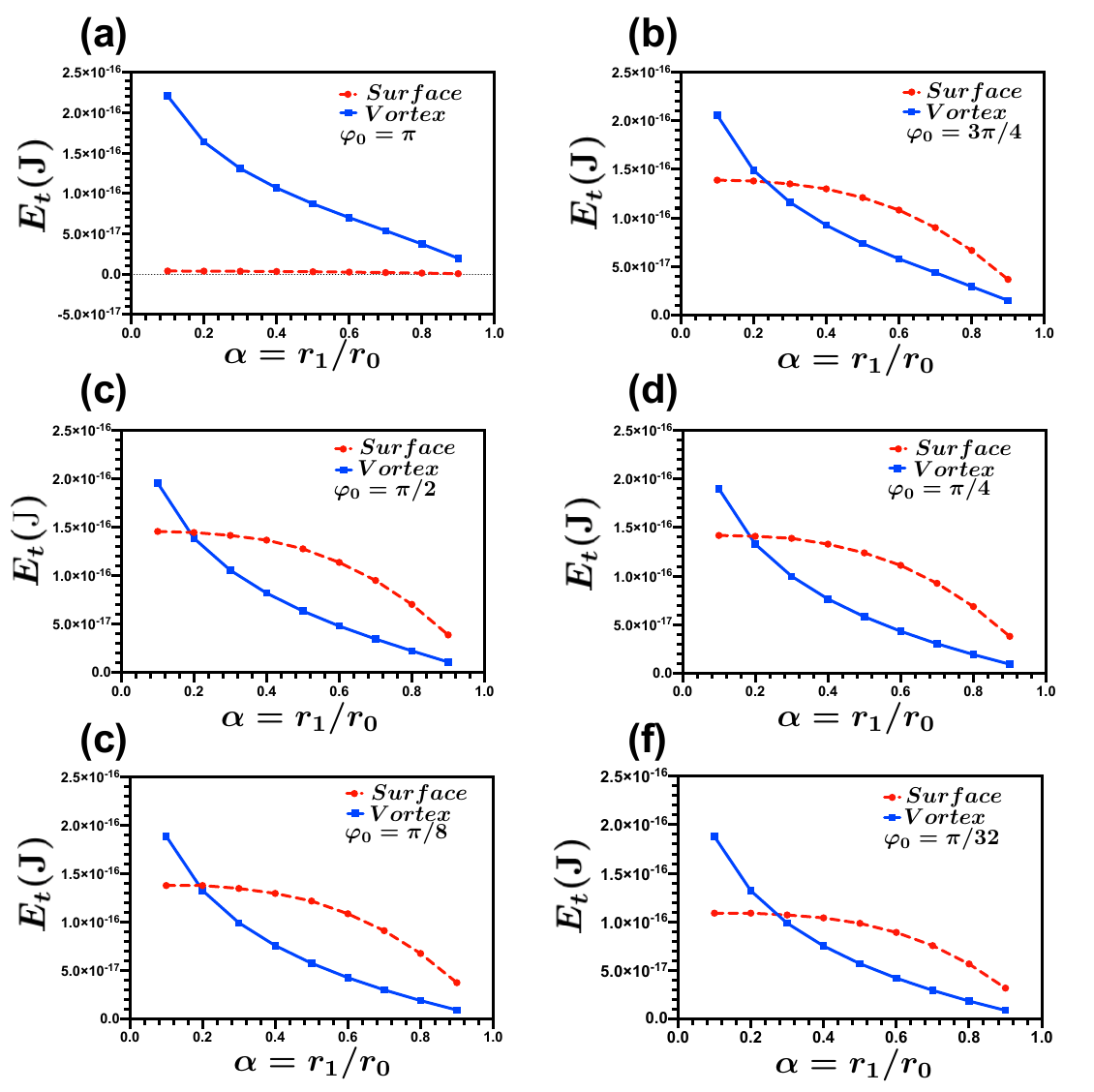}\caption{Total magnetic energy of the $IS$ and  $V$ states as a function of the BNT thickness $\alpha$ for (a) $\varphi_0=\pi$, (b) $\varphi_0=3\pi/4$, (c) $\varphi_0=\pi/2$, (d) $\varphi_0=\pi/4$, (e) $\varphi_0=\pi/8$, and (f) $\varphi_0=\pi/32$. The nanotube has a length and external radius of $L=1000$ nm and $r_0=50$ nm, respectively.}\label{Energy_curve}
\end{figure}

WE start calculating the energies as a function of the internal and external radii  and the curvature of the BNT for each considered magnetization configuration. All results were obtained  using permalloy parameters, that is, a stiffness constant $A=1.3\times10^{-11}$ J/m and a saturation magnetization  $M_s=860$ kA/m. Figure  \ref{Energy_curve}  shows the total magnetic energy as a function of the aspect ratio $\alpha=r_1/r_0$, for a BNT with $L=1000$ nm, $r_0=50$ nm, and different curvatures (from Fig. \ref{Energy_curve}(a) for a torus to Fig. \ref{Energy_curve}(f) that corresponds to an almost straight NT). From these figures we evidence that the magnetic ground state depends on both, the thickness of the tube  (related to $\alpha$) and its curvature (related to $\varphi_0$). Fig. \ref{Energy_curve}(a) shows that the torus configuration, when $\varphi_0=\pi$, presents an $IS$ configuration for all thickness. This result is in agreement with Ref. \cite{Allison-JMMM-2019}, in which the authors analyzed the energetics of tangential magnetic states in toroidal shells. From Figs. \ref{Energy_curve}(b)-(f), we observe that there is a critical thickness, $\alpha_c$, at which both magnetic configurations have the same magnetic energy. For $\alpha <\alpha_c$, the $IS$ state appears as the magnetic ground state for the BNT, while for $\alpha >\alpha_c$ the $V$ state is the lower energy configuration. Fig. \ref{Energy_curve}(f) exhibits a good agreement with results for  straight nanotube presented by Escrig {\it al.}\cite{Escrig-JMMM-2007}.

\begin{figure}
\includegraphics[scale=0.5]{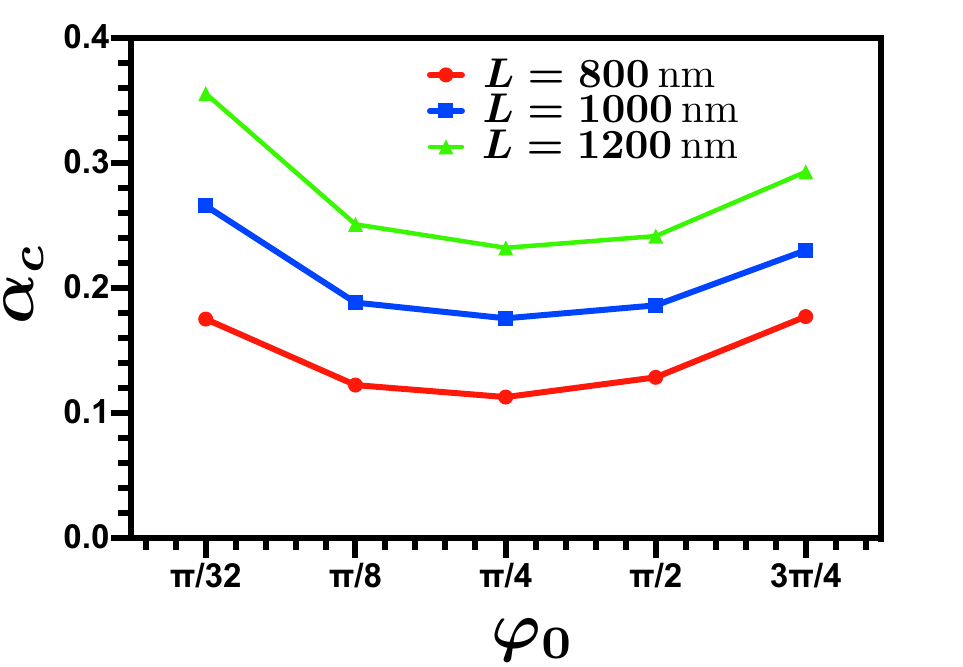}\caption{Estimated value of $\alpha_c$ for the transition of the $IS$ configuration to the $V$ state as a function of the curvature of the BNT, and for different lenghts.}\label{betacritico}
\end{figure}

To see more clearly the behavior of the critical thickness $\alpha_c$,  we summarize $\alpha_c$ obtained from Fig. \ref{Energy_curve}(b)-(f) in Fig. \ref{betacritico} for different values of $L$.
Figure  \ref{betacritico} illustrates our results as a function of the curvature, $\varphi_0$, for different values of $L$. Two different behaviors can be observed in this figure. From one side,  a monotonic behavior of  $\alpha_c$ as a function of $L$, i.e., the increase in the BNT length leads to an increase in $\alpha_c$. On the other hand, the increase in the curvature produces a non-monotonic behavior of $\alpha_c$. In fact, it can be observed that $\alpha_c$ decreases when we increase the curvature for low curvature. However, when $\varphi_0 \approx \pi/4$, $\alpha_c$ starts increasing when $\varphi_0$ increases for $ \pi/4 \lessapprox \varphi_0 < \pi$.

\begin{figure}
\includegraphics[scale=0.7]{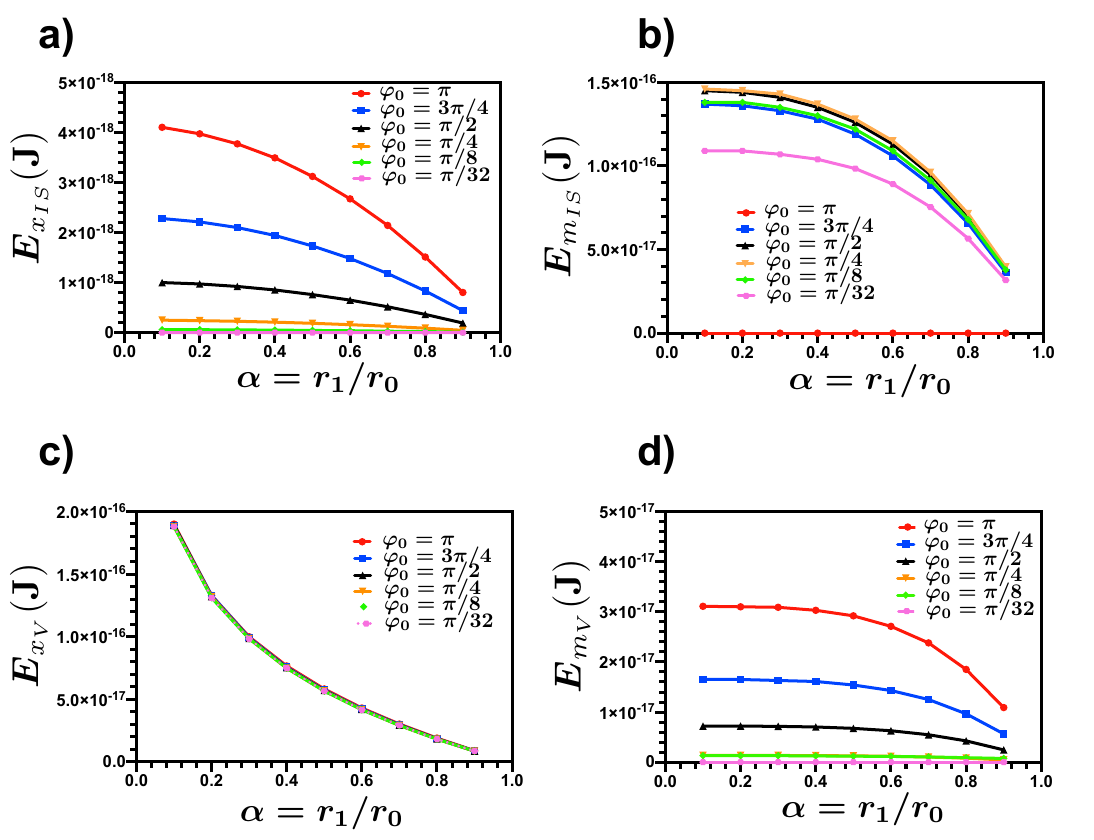}\caption{Energy terms for  $IS$ and $V$ states. (a) and (b) illustrate the exchange and magnetostatic energies of the $IS$ state, respectively. In (c) and (d) we present the exchange and magnetostatic energies of  the $V$ state, respectively.}\label{Energy_Analysis}
\end{figure}

To understand the non-monotonic behavior of $\alpha_c$ as a function of $\varphi_0$  (Fig. \ref{betacritico}) and also the dependence of the ground state as a function of $\alpha$ (Fig. \ref{Energy_curve}), we need to analyze the exchange and magnetostatic energies for the $IS$ and $V$ states separately. Figure \ref{Energy_Analysis} shows the  exchange and magnetostatic energies as a function of $\alpha$ for different values of $\varphi_0$, for a BNT with $L=1000$ nm and $r_0=50$ nm. Figure \ref{Energy_Analysis}(a) shows the exchange energy of the $IS$ state, where this energy depicts a monotonic behavior when varying $\alpha$ and $\varphi_0$.   Figure \ref{Energy_Analysis}(b) illustrates  the magnetostatic energy for the $IS$ configuration, where we observe a monotonic behavior with $\alpha$, but a non-monotonic behavior with $\varphi_0$. Indeed, we observe an abrupt drop of the magnetostatic energy to zero for the torus geometry, $\varphi_0=\pi$. This non-monotonic behavior of the magnetostatic energy is due to the term $\sin^2(k\varphi_0)$ in Eq. \eqref{dipEneIS}. Figure \ref{Energy_Analysis}(c) shows the exchange energy of the $V$ state, where we observe a monotonic behavior with $\alpha$. Nevertheless, it can be notice that the obtained values for the exchange energy in the $V$ state are almost invariant to changes in $\varphi_0$. This behavior occurs  because the value of the first term in Eq. \eqref{exEneV} is predominant, and it is practically constant.  Finally, Figure \ref{Energy_Analysis}(d) shows the magnetostatic energy of the $V$ state. In this case, we also observe a monotonic behavior of the magnetostatic contribution to the energy with $\alpha$ and $\varphi_0$. Therefore, the only contribution that has non-monotonic behavior is the magnetostatic energy obtained for the $IS$ state, and then this term is the responsible by the non-monotonic behavior for the critical thickness, $\alpha_c$. Indeed, the maximum value of the magnetostatic energy for the $IS$ state when we change $\varphi_0$ is at $\varphi_0=\pi/4$, i.e., the same minimum observed in Fig. \ref{betacritico}.

In addition, from Fig. \ref{Energy_Analysis}, we can understand the dependence of the ground state as a function of $\alpha$. We observe that due to the high exchange energy of the $V$ state compared to the other energy contributions for smaller values of $\alpha$ ($E_{x_V}\gtrsim 1.5 \times 10^{-16}$ J), the $IS$ state appears as the magnetic ground state for BNT for $\alpha<\alpha_c$. On the other hand, when $\alpha\gtrsim0.2\sim \alpha_c$, there is a reduction of the exchange and magnetostatic energy of both configurations, see Figs. \ref{Energy_Analysis}(a)-(d). However, the magnetostatic energy of the $IS$ configuration starts to have higher energy than the other energies when we increase $\alpha$, e.g., for $\alpha \approx 0.8$, the higher energy is the magnetostatic energy of the $IS$ state $E_{m_{IS}}\approx  5\times 10^{-17}$ J, see Fig. \ref{Energy_Analysis}(b). Therefore, the competition between magnetostatic and exchange energies make the vortex state more favorable when $\alpha$ increases. In addition, Fig. \ref{Energy_curve}(a) presents the energy for a toroidal geometry where the $IS$ configuration is observed for all $\alpha$ values, due to the low exchange energy  (of the order of $10^{-18}$ J) and the absence of the magnetostatic energy for $\varphi_0=\pi$ ($E_{m_{IS}}=0$) (see Figs. \ref{Energy_Analysis}(a) and \ref{Energy_Analysis}(b), respectively).

\section{Conclusions}\label{ThConclusions}
Analytical expressions for the magnetic energy of different magnetization configurations have been obtained by representing a bent nanotube as a torus section. By using a curvilinear basis, exchange and magnetostatic energy for the vortex and the in-surface magnetization configurations have been explicitly calculated. Our results evidence a dependence of the energy on the curvature of the BNT. The numerical analysis of the obtained equations reveals that the magnetization ground state depends on the curvature, the thickness, and the length of the BNT. We have also obtained the value of the critical thickness for which occurs the transition from the $IS$ to the $V$ state. For  thick tubes, we observe the $IS$ state, and for thin structures, the vortex configuration is the lower energy state. This behavior can be explained looking to the exchange energy for the $V$ state (magnetostatic energy for the $IS$ state) which exhibits  a dominant contribution for thick (thin) structures. In addition, we observed a non-monotonic behavior as a function of  curvature, of the critical thickness at which the transition between both states appears . These results could be guidelines for the development of spintronic devices.

\section*{Acknowledgements}
We acknowledge financial support in Chile from FONDECYT Grants 1161018 and 1160198, and Financiamiento Basal para Centros Cient\'{i}ficos y Tecnol\'{o}gicos de Excelencia FB 0807.  D.M.-A. acknowledges Postdoctorado FONDECYT 2018, folio 3180416. M.A.C. acknowledges Conicyt-PCHA/Doctorado Nacional/2017-21171016. In Brazil, we thank the financial support from CNPq (Grants 401132/2016-1 and 309484/2018-9).

\appendix

\section{Details on the magnetostatic potential calculations}\label{Apendice}

\noindent

In the absence of electric currents, the magnetostatic potential comes from surface, $\sigma=\mathbf{m}\cdot\mathbf{n}$, and volumetric,  $\chi=- \boldsymbol {\nabla}\cdot\mathbf{m}$, magnetic charges, where  $\mathbf{n}$ is the normal vector pointing outward the surface of the structure. The magnetostatic potential  is
\begin{eqnarray}
\Phi(\mathbf{r}')\equiv\Phi=\frac{M_s}{4\pi}\left[\int_S\frac{\sigma}{\mathcal{R}}dS+\int_V\frac{\chi}{\mathcal{R}}dV\right]\,,
\end{eqnarray}

\noindent
where $\mathcal{R}\equiv|\mathbf{r}\,'-\mathbf{r}|$ is the distance between two points inside the magnetic body, and it can be written as 
\begin{eqnarray}\label{GreenFunc}
\frac{1}{\mathcal{R}}=\frac{\sqrt{\Upsilon'\, \Upsilon}}{\pi\,\rho}\sum_{k,n=0}^{\infty}(-1)^k\varepsilon_n\varepsilon_k\cos [n(\eta'-\eta)]\cos [k(\varphi'-\varphi)]\nonumber\\\times\frac{\Gamma(n-k+1/2)}{\Gamma(n+k+1/2)}P_{n-1/2}^k(\cosh\beta_<)Q_{n-1/2}^k(\cosh\beta_>)\,,\,\,\,
\end{eqnarray}
If we  use the property \cite{Gradshtein-Book} 
\begin{eqnarray}
\frac{\Gamma(n-k+1/2)}{\Gamma(n+k+1/2)}Q_{\nu}^{\tau\,\,\,\,x}=Q_{\nu}^{-\tau\,\,\,\,x}\,.
\end{eqnarray}
and  also  the next notation
\begin{eqnarray}
P^{\tau}_{\nu}(\cosh x)=P^{\tau\,\,\,\,x}_{\nu}\,\,\,\,\,\,\,\text{and}\,\,\,\,\,\,\,
Q^{\tau}_{\nu}(\cosh x)=Q^{\tau\,\,\,\,x}_{\nu}\,.
\end{eqnarray}
we can write
\begin{eqnarray}\label{GreenFunc}
\frac{1}{\mathcal{R}}=\frac{\sqrt{\Upsilon'\, \Upsilon}}{\pi\,\rho}\sum_{k,n=0}^{\infty}(-1)^k\varepsilon_n\varepsilon_k\cos [n(\eta'-\eta)]\cos [k(\varphi'-\varphi)] P_{n-1/2}^{k\,\,\,\,\beta_<}\,Q_{n-1/2}^{-k\,\,\,\,\beta_>}\,,\,\,\,
\end{eqnarray}

\subsection{Vortex state}
Due to the absence of surface magnetic charges, the $V$ state presents a dipolar energy related to volumetric magnetostatic charges. In simple toroidal coordinates, the magnetization of the vortex state is $\mathbf{m}_{{{V}}}=\boldsymbol{\hat{\theta}}$.  The volumetric charge is given by $\chi=- \boldsymbol {\nabla}\cdot\mathbf{m}=-\frac{1}{h_r h_{\phi} h_{\theta}}\frac{\partial (h_r h_\phi)}{\partial \theta}$, where $h_r=1$, $h_\phi=R+r\sin \theta$, and $h_\phi=r$ are the scale factors.
Then, we obtain $\chi=-\frac{\cos \theta}{R+r\sin \theta}=-\frac{1}{r_0}\frac{\sin \eta}{\sinh \beta} \frac{\sqrt{\cosh (\beta)-\cos (\eta )}}{  \sqrt{\cosh (\beta -2 \text{$\beta_0 $})-\cos (\eta)}}$. Therefore, the magnetostatic potential related to this configuration is given by
\ba
\tilde\Phi_{V}=-\frac{M_S}{4\pi}\int_{V}  \frac{1}{r_0 \mathcal{R} }\frac{\sin \eta}{\sinh \beta} \frac{\sqrt{\cosh (\beta)-\cos (\eta )}}{  \sqrt{\cosh (\beta -2 \text{$\beta_0 $})-\cos (\eta)}} dV\,.
\ea
\noindent
From expanding the inverse of the distance in the form of Eq.~(\ref{GreenFunc}) and taking the volume element given by $dV= \frac{\rho ^3 \sinh (\beta )}{(\cosh (\beta )-\cos (\eta ))^3} d\beta d\eta d\phi$, the above equation can be rewritten as

\ba
\tilde\Phi_{V}=-\frac{M_S \rho_0^2}{4\pi^2 r_0} \sqrt{\Upsilon'\, } \int_{}  d\beta d\eta   \sum_{k,n=0}^{\infty} \frac{ (-1)^k \varepsilon_n\varepsilon_k \sin \eta \cos [n(\eta'-\eta)] }{ [\cosh (\beta -2 \text{$\beta_0 $})-\cos (\eta)]^{1/2} \Upsilon^{2} } h_k(\varphi')  P_{n-1/2}^{k\,\,\,\,\beta_<}\,Q_{n-1/2}^{-k\,\,\,\,\beta_>} 
 \,.
\ea
\noindent
where $h_k(\varphi')=k^{-1}[2\cos ( k\varphi')\sin( k\varphi_0)]$.
The integral in $\eta$ can be performed  using the trigonometric identity $\cos [n(\eta'-\eta)]=\cos (n\eta)\cos (n\eta')+\sin (n\eta)\sin (n\eta')$ 

\ba
\tilde\Phi_{V}=-\frac{M_S \rho_0^2}{4\pi^2 r_0} \sqrt{\Upsilon'\, } \int_{}  d\beta d\eta   \sum_{k,n=0}^{\infty} \frac{ (-1)^k \varepsilon_n\varepsilon_k \sin \eta  \sin (n\eta)\sin (n\eta')}{ [\cosh (\beta -2 \text{$\beta_0 $})-\cos (\eta)]^{1/2} \Upsilon^{2} } h_k(\varphi')  P_{n-1/2}^{k\,\,\,\,\beta_<}\,Q_{n-1/2}^{-k\,\,\,\,\beta_>} 
 \,.
\ea
\subsection{In-surface configuration}
The magnetostatic potential of the $IS$ state is determined by the the magnetic surface charges in the borders of the BNT, given by 
\begin{eqnarray}\label{Potx11}
\tilde\Phi_{{\text{IS}}}=\frac{M_S}{2\pi}\int_{1} \frac{1}{\mathcal{R}_1}dS_{1} -\frac{M_S}{2\pi}\int_{1} \frac{1}{\mathcal{R}_2}dS_{2} \,,
\end{eqnarray}
where $dS_{1}=dS_{2}=\rho_0^2/(\cosh\beta-\cos\eta)^2\,d\eta\, d\beta$ and
\begin{eqnarray}\label{GreenFunc1}
\frac{1}{\mathcal{R}_1}=\frac{\sqrt{\Upsilon'\, \Upsilon}}{\pi\,\rho}\sum_{k,n=0}^{\infty}(-1)^k\varepsilon_n\varepsilon_k\cos [n(\eta'-\eta)]\cos [k(\varphi'-\varphi_0)] P_{n-1/2}^{k\,\,\,\,\beta_<}\,Q_{n-1/2}^{-k\,\,\,\,\beta_>}\,,\,\,\,\\
\label{GreenFunc2}
\frac{1}{\mathcal{R}_2}=\frac{\sqrt{\Upsilon'\, \Upsilon}}{\pi\,\rho}\sum_{k,n=0}^{\infty}(-1)^k\varepsilon_n\varepsilon_k\cos [n(\eta'-\eta)]\cos [k(\varphi'+\varphi_0)] P_{n-1/2}^{k\,\,\,\,\beta_<}\,Q_{n-1/2}^{-k\,\,\,\,\beta_>}\,,\,\,\,
\end{eqnarray}

The substitution of Eqs.~(\ref{GreenFunc1})  and (\ref{GreenFunc2}) in Eq. (\ref{Potx11}) yields
\begin{eqnarray}
\tilde\Phi_{IS}=-\frac{M_S\rho_0\,\sqrt{\Upsilon'}}{{2\pi^2}}\int d\beta \,d\eta\sum_{k,n=0}^{\infty}(-1)^k\varepsilon_n\varepsilon_k\, \frac{\cos[n(\eta-\eta')]\, }{\Upsilon^{3/2}} g_k(\varphi')P_{n-1/2}^{k\,\,\,\,\beta_<}\,Q_{n-1/2}^{-k\,\,\,\,\beta_>}\,,\end{eqnarray}
where $g_k(\varphi')=\sin (k\varphi_0)\,\sin (k\varphi')$. In the same way as in previous calculations, the integrals in $\eta$ can be performed  using the trigonometric identity $\cos [n(\eta'-\eta)]=\cos (n\eta)\cos (n\eta')+\sin (n\eta)\sin (n\eta')$ and the integral representation of $Q_{n-1/2}^k(\cosh\beta)$ (see Ref. \cite{Gradshtein-Book}). In this case, after some algebraic manipulation, we obtain Eq. \eqref{PotSX1}.


\begin{thebibliography}{99}
\bibitem{Paper-1}
R. Hertel, Nat. Nanotechnol. 8, 318 (2013).

\bibitem{paper-3}
A.P. Guimar\~aes, \textit{Principles of Nanomagnetism}, (Springer, Heidelberg, 2009).

\bibitem{paper-4}
R. Streubel, D.J. Thurmer, D. Makarov, F. Kronast, T. Kosub, V.P. Kravchuk, D.D.
Sheka, Y. Gaididei, R. Sch√§fer, and O.G. Schmidt, Nano Lett. 12, 3961 (2012).

\bibitem{KaiLiu-paper}
D.W. Shi, P.K. Greene, P. Liu, K. Javed, K. Liu, and X. F. Han, IEEE Trans. Magnetics \textbf{50}, 2303004 (2014).

\bibitem{Knielsh-paper}
I. Minguez-Bacho, S. Rodriguez-L\'opez, M. V\'azquez, M. Hern\'andez-V\'elez, and K. Nielsch, Nanotechnology 25, 145301 (2014).

\bibitem{experimental-techniques}
M. V\'azquez, in \textit{Advanced Magnetic Nanowires} ed. by H. Krom\"uller, and S. Parkin, Handbook of Magnetism and Advanced Magnetic Materials, vol. 4 (Wiley, Chichester, 2007).

\bibitem{Gaididei-PRL-2014}
Y. Gaididei, V.P. Kravchuk, and D.D. Sheka, Phys. Rev. Lett. \textbf{112}, 257203 (2014).

\bibitem{Review-Curvature}
R. Streubel, P. Fischer, F. Kronast, V.P. Kravchuk, D.D. Sheka, Y. Gaididei,
O.G. Schmidt, and D. Makarov, J. Phys. \textbf{D 49}, 363001  (2016).

\bibitem{Review-3DNanomagnetism}
A. Fern\'andez-Pacheco, R. Streubel, O. Fruchart, R. Hertel, P. Fischer, and R.P. Cowburn, Nat. Commun. \textbf{8}, 15756 (2017). 

\bibitem{Landeros-JAP-2010}
P. Landeros and \'A. S. N\'u\~nez, J. Appl. Phys. \textbf{108}, 033917 (2010).

\bibitem{Hertel-Spin-2013}
R. Hertel, SPIN 3, 1340009 (2013).

\bibitem{Pylypvskyi-SRep-2016}
O.V. Pylypovskyi, D.D. sheka, V.P. Kravchuk, K.V. Yershov, D. Makarov, and Y. Gaididei, Sci. Rep. \textbf{6}, 23316 (2016).

\bibitem{Smiljan-JAP-2017}
S. Vojkovic, V.L. Carvalho-Santos, J.M. Fonseca, and A.S. Nunez, J. Appl. Phys. \textbf{121}, 113906 (2017).

\bibitem{Otalora-PRL-2017}
J. A. Ot\'alora, M. Yan, H. Schultheiss, R. Hertel, and A. K\'akay, Phys. Rev. Lett. \textbf{117}, 227203 (2016).  


\bibitem{dots1}
J.-M. Hu, T. Yang, K. Momeni, X. Cheng, L. Chen, S. Lei, S. Zhang, S.
Trolier-McKinstry, V. Gopalan, G. P. Carman, C.-W. Nan, and L. Q.    
Chen, Nano Lett. \textbf{16}, 2341 (2016).



\bibitem{rings1}
M. Goiriena-Goikoetxea, K.Y. Guslienko, M. Rouco, I. Orue, E. Berganza, M. Jaafar, A. Asenjo, M.L. Fern\'andez-Gubieda, L. Fern\'andez Barqu\'in, and A. Garc\'ia-Arribas, Nanoscale \textbf{9}, 11269 (2017). 

\bibitem{wires}
P. Landeros, S. Allende, J. Escrig, E. Salcedo, D. Altbir, and E. E. Vogel, Appl. Phys. Lett. \textbf{90}, 102501 (2007).

\bibitem{Knielsh}
K. Nielsch, F. J. Casta\~no, S. Matthias, W. Lee, and C. Ross, Adv. Eng. Mater. \textbf{7}, 217 (2005).   

\bibitem{Jaworowicz-Nanotechnology-2009}
J. jaworowicz, N. Vernier, J. Ferr\'e, A. Maziewski, D. Stanescu, D. Ravelosona, A.S. Jacqueline, C.Chappert, B. Rodmacq, and B. Di\'eny, Nanotechnology \textbf{20}, 215401 (2009).  

\bibitem{Allwood-Science-2005}
D. A. Allwood, G. Xiong, C. C. Faulkner, D. Atkinson, D. Petit, and R.P. Cowburn, Science \textbf{309}, 1688 (2005).   

\bibitem{Son-JChem}
S. J. Son, J. Reichel, B. He, M. Schuchman, and S. B. Lee, J. Am. Chem. Soc. \textbf{127}, 7316 (2005).


\bibitem{Parkin-Science-2008}
S.S.P. Parkin, M. Hayashi, and L. Thomas, Science \textbf{320}, 190 (2008).


\bibitem{theoretical-studies}
Y.P. Ivanov, M. V\'azquez, and O.Chubykalo-Fesenko, J. Phys. D. \textbf{46}, 485001 (2013).

\bibitem{Yan-paper}
M. Yan, A. K\'akay, S. Gliga, and R. hertel, Phys. Rev. Lett. \textbf{104}, 057201 (2010).

\bibitem{theoretical-studies2}
R. Wieser, E.Y. Vedmedenko, P. Weinberger, and R. Wiesendanger, Phys. Rev. \textbf{B 82}, 144430 (2010).

\bibitem{theoretical-studies3}
R. Hertel, and A. K\'akay, J. Magn. Magn. Mat. \textbf{379}, 45 (2015).

\bibitem{Walker-paper}
N.L. Schryer, and L.R. Walker, J. Appl. Phys. \textbf{45}, 5406 (1974).

\bibitem{Mougin-paper}
A. Mougin, M. Cormier, J.P. Adam, P.J. Metaxas, and J. Ferr\'e, Europhys. Lett. \textbf{78}, 57007 (2007).


\bibitem{Kravchuk-wire-paper}
K.V. Yershov, V.P. Kravchuk, D.D. Sheka, and Y. Gaididei, Phys. Rev. \textbf{B 92}, 104412 (2015).

\bibitem{Moreno-PRB-2017}
R. Moreno, V.L. Carvalho-Santos, A.P. Espejo, D. Laroze, O. Chubykalo-Fesenko, and D. Altbir, Phys. Rev. \textbf{B 96}, 184401 (2017).


\bibitem{Escrig-JMMM-2007}
J. Escrig, P. Landeros, D. Altbir, E.E. Vogel, and P. Vargas, J. Magn. Magn. Mat. \textbf{308}, 233 (2007).



\bibitem{Morse-Book}
P. M. Morse, and H. Feshbach, \textit{Methods of Theoretical Physics}, (MCGraw-Hill,  
New York, 1953).

\bibitem{Gradshtein-Book}
I.S. Gradshteyn, and I.M. Ryzhik, \textit{Table of Integrals, Series and Products},   
7th ed. (Academic, New York, 2007).

\bibitem{Allison-JMMM-2019}
A.W. Teixeira, S. Castillo-Sep\'ulveda, S.Vojkovic, J.M.Fonseca, D.Altbir, \'A.S. N\'u\~nez, and V.L.Carvalho-Santos, J. Magn. Magn. Mat. \textbf{478}, 253 (2019). 









\end{thebibliography}
\end{document}